\documentclass[12pt]{article}
\def\e{{\rm e}}

\title{\bf Decorating Random Quadrangulations}
\author{ 
{\em 
Desmond A. Johnston}\\ 
Dept. of Mathematics\\
Heriot-Watt University\\
Riccarton\\
Edinburgh, EH14 4AS, Scotland\\
\\
{\bf and}\\
\\
{\em Ranasinghe P. K. C. Malmini}\\
Department of Mathematics\\
University of Sri Jayewardenepura\\
Gangodawila, Sri Lanka.} 
\begin{document}
  \maketitle
{\Large
                      \begin{abstract}
On various regular lattices (simple cubic, body centred cubic..)
decorating an edge with an  Ising spin coupled 
by bonds of strength $L$ to the original vertex spins
and competing with a direct anti-ferromagnetic bond
of strength $\alpha L$  can give rise to three transition
temperatures for suitable $\alpha$.
The system passes through ferromagnetic, paramagnetic,
anti-ferromagnetic and paramagnetic phases respectively as the temperature is
increased.

For the square lattice on the other hand 
multiple decoration is required to see this effect.
We note here that a single decoration suffices
for the Ising model on planar random quadrangulations
(coupled to 2D quantum gravity). Other random bipartite
lattices such as the Penrose tiling are more akin to
the regular square lattice and require multiple decoration
to have any affect.

\end{abstract}
}
\thispagestyle{empty}
\newpage
%
%
The application of various classes of transformations,
such as duality, star-triangle, and decoration-iteration
to the Ising model on regular two dimensional lattices has been exhaustively
investigated in the past \cite{syozi} both as a means of obtaining
solutions on new lattices and as a way of modelling the physical
properties of substances with more complicated behaviour than simple
ferromagnetism. The decoration-iteration transformation acts
exclusively at the level of the bonds in the lattice, so it 
knows nothing about the larger scale structure. It is
therefore still applicable on lattices which exhibit some
form of geometrical disorder.
In this short note we look at the effect of a decoration-iteration 
transformation on an Ising model on just such a 
class of lattices. We consider a 
model with competing
ferromagnetic and anti-ferromagnetic interactions living on an ensemble
of random quadrangulations (i.e. coupled to two-dimensional 
quantum gravity).

The basic decoration-iteration transform is shown in Fig.1,
where summing over the central spin $s$ with
couplings $L$ gives rise to a 
new effective coupling $K$ between the 
primary vertex spins $\sigma_1, \sigma_2$
\begin{equation}
\sum_{s} \exp \left[ L s ( \sigma_1 + \sigma_2 ) \right] = A \exp ( K \sigma_1 \sigma_2),
\end{equation}
where $A =2 ( \cosh ( 2 L ) )^{1/2}$ and $\exp ( 2 K ) = \cosh ( 2 L )$, i.e.
\begin{equation}
K = \left( \frac{1}{2} \right) \log \left[ \cosh ( 2 L ) \right].
\label{e2}
\end{equation}
If a direct anti-ferromagnetic bond is stirred into the mix as well
for good measure this becomes
\begin{equation}
K =  -\alpha L + \left( \frac{1}{2} \right) \log \left[ \cosh ( 2 L ) \right].
\label{e2a}
\end{equation}
This may easily be extended to a situation such as that shown in Fig.2 where
we  introduce $n$ decorating spins as well as the direct anti-ferromagnetic bond.
In this case summing over the intermediate spins
gives
\begin{equation}
K  = - \alpha L +  \left( \frac{1}{2} \right) \log \left[ {( \exp( 2 L ) + 1 )^{n+1}
+ ( \exp( 2 L ) -1 )^{n+1} \over ( \exp( 2 L ) + 1 )^{n+1}
- ( \exp( 2 L ) -1 )^{n+1} } \right]  
\label{en}
\end{equation}
along with an equation for the normalisation factor
\begin{equation}
A^2 = 2^n {( \sinh ( 2 L ))^{n+1}\over \sinh ( 2 K ) }.
\end{equation}

It was pointed out by Nakano \cite{nakano} that generically the form of the transformation
in both equs.(\ref{e2a},\ref{en}) meant that {\it three} transitions could
occur for an Ising model. 
Whether this behaviour actually occurred or not depended on the 
critical temperature values for a given lattice, along with 
the value of $\alpha$ and the degree $n$
of iteration. 
For the singly decorated model,
the minimum value of $K$ in equ.(\ref{e2a}), for instance,
is $K_{min} = -(1/2) \log ( 2) \sim -0.3196..$ which is attained as $\alpha \to 1$.
Since this is larger than the critical value of the coupling
at the anti-ferromagnetic transition on the square lattice
$K_{crit} = -0.44609..$ the $K(L)$ curve can at best
intersect the ferromagnetic transition value $K_{crit} = + 0.44609..$ and only one
(ferromagnetic) transition will be in evidence in the decorated model.
A generic curve displaying this behaviour is shown in Fig.3, which is
for $\alpha=4/5$. We can see that $K(L)$ does not dip below the line
at $K_{crit}= -0.44609..$, but does cross  $K_{crit} = + 0.44609..$. As $\alpha$
is increased still further $K(L)$ eventually becomes monotone decreasing, 
attaining $K_{min}$ as $\alpha \to 1$, so even the 
ferromagnetic transition disappears.

This behaviour is strongly lattice dependent since we are looking 
for intersections of the $K(L)$ curve with the critical values of
the coupling. Since $K_{crit} = \pm 0.2217..$ for the simple cubic lattice
and $K_{crit}= \pm 0.1574$ for the body centred cubic lattice, these will
display multiple transitions when the $K(L)$ curve cuts through the
anti-ferromagnetic critical coupling values for sufficiently large
$\alpha$.
As $L$ is decreased (i.e. the temperature is increased) following the
$K(L)$ curve then takes one from a ferromagnetic phase via a
paramagnetic phase to an anti-ferromagnetic phase and finally
to a paramagnetic phase again. One thus has a sequence
of one ferromagnetic and two anti-ferromagnetic transitions.

In short, for a lattice with an anti-ferromagnetic critical coupling
of sufficiently small modulus, $|K_{crit}| < |K_{min}|$, 
multiple transitions are to be expected 
when competing direct and decorated bonds are present. For
$n$-fold decoration the minimum value of $K(L)$ tends to
$K_{min,~n} = - ( 1 /2 ) \log ( n +1 )$ as $\alpha \to 1$, so $n=2$ is
sufficient to induce a triple transition in even the square lattice
Ising model.

We now turn to the case of the Ising model
coupled to 2D quantum gravity as an example
of the application of decoration-iteration transformations
on geometrically disordered lattices. 
The partition function for the Ising model on a single planar graph
$G^n$ with $n$
vertices is just \cite{BK}
\begin{equation}
Z_{{\rm single}}(G^n,K) =
\sum_{\{\sigma\}} \e^{{K}\sum_{<i,j>} \sigma_i
\sigma_j}\; .
\end{equation}
The coupling to gravity is incorporated by introducing a 
sum   
over some class of planar graphs ${\{G^n\}}$ 
\begin{equation}
Z_n(K) = \sum_{\{G^n\}} Z_{{\rm single}}(G^n,K)\, .
\label{sum}
\end{equation}
The grand canonical partition function for this model
\begin{equation}
{\cal Z} = \sum_{n=1}^{\infty} \left( - 4 g c \over ( 1 - c^2)^2 \right)^n
Z_n(K)
\label{grand}
\end{equation}
where $c = \exp ( - 2 K)$
can be expressed as the free energy
\begin{equation}
{\cal Z} = - \frac{1}{N^2}\log \int {\cal D}\phi_1~{\cal D}\phi_2~ \exp~\left( -Tr[{1\over 2}(\phi_1^2+\phi_2^2)-c\phi_1\phi_2  - \frac{g}{4}( \phi_1^4 + \phi_2^4)]  \right)
\label{matint}
\end{equation}
of a matrix model, 
where we have specifically given the potential which generates
$\phi^4$ graphs and
$\phi_{1,2}$ are $N \times N$ Hermitian matrices.
The $N \to \infty$ limit is taken to pick out planar graphs.

The model as defined above has spins
living on the vertices of $\phi^4$ planar graphs and 
has a critical coupling of $c_{crit}=1/4$, $K_{crit} = 0.69314..$
\cite{BK}.
It displays
no anti-ferromagnetic transition, because both odd and even loops 
are present. The decoration process outlined above will thus not
induce multiple transitions  since there is no
anti-ferromagnetic line for the curve $K(L)$ to cross. It is a similar
story for both $\phi^3$ planar graphs and their dual triangulations.

However the dual to the $\phi^4$ graphs, random quadrangulations, 
are rather more
amenable to decoration. We can see from Fig.4 that, although
the number of squares round a vertex is arbitrary, the fact that every face
is a square means that the random quadrangulation is bipartite
and hence would be expected to have  an anti-ferromagnetic transition.
This has been confirmed both by direct simulation \cite{DJ}
and by studies
of the Fisher (temperature) zeroes of the partition function
\cite{jan,JJS}.
The critical coupling for the ferromagnetic transition on the
random quadrangulations is given by the dual of the $\phi^4$ value,
namely $ c^*_{crit} = ( 1 - c_{crit} ) / ( 1 + c_{crit} ) = 3/5$ and the 
anti-ferromagnetic value by its inverse, $5/3$. If we translate
these back to $K$ we find 
$K_{crit} = \mp (1/2) \log ( 3 / 5)
\sim \pm 0.255412..$ for the ferromagnetic
and anti-ferromagnetic critical
couplings respectively.
We can immediately see that since $|K_{crit}| < |K_{min}|$
it is now possible for the curve for a singly decorated $K(L)$ to
cross the anti-ferromagnetic line for a suitable $\alpha$ value. In Fig.5 we
show $K(L)$ for $\alpha = 0.92$, where the three transition points
may be seen. All of the transitions
will display the $KPZ$ \cite{kpz,ddk}
exponents $\alpha = -1, \beta = 1/2, \gamma=2$.
Thus, unlike the regular square lattice, the Ising model 
with singly decorated and competing bonds on random
quadrangulations will display a sequence of one ferromagnetic
and two anti-ferromagnetic transitions as the temperature is increased.

As we have noted, the decoration results are lattice dependent
since they deal with critical temperatures. The ensemble of random
quadrangulations we discussed above includes degenerate gluings
of the squares along multiple edges, since the original $\phi^4$
graphs do not exclude self-energy and vertex correction diagrams.
No analytical results are available for an ensemble of
``regular'' random quadrangulations, where no multiple gluings
are allowed, but the simulation of \cite{DJ} gave an
estimate of $K_{crit} \sim 0.4$, which lies outside the
range for multiple transitions
with a single decoration. In this case we would 
{\it not} expect to see the sequence of transitions discussed above.
However,
doubly decorated bonds have a minimum of $K(L)$ at
$K_{min,~2} \sim -0.5493...$ and thus would give rise
to multiple transitions for these regular random quadrangulations,
just as they do for the regular square lattice Ising model.

It is also possible, of course, to elaborate the decoration procedure
in various ways \cite{fish}. Introducing a higher spin $s$ which can take values
$-S,\, -S+1, \ldots, S-1, S$ as the decorating spin
and taking its interaction with the Ising spins to be
\begin{equation}
E = - \frac{L}{S} s \sigma
\end{equation}
gives an effective coupling
\begin{equation}
K = -\alpha L + \frac{1}{2} \log \left[ {\sinh \left( { (2 S + 1 ) L \over S} \right)
\over ( 2 S + 1) \sinh \left( { L \over S} \right) } \right]
\end{equation}
(where we have again allowed for a direct anti-ferromagnetic bond). For this higher
spin decoration $K(L)$ still looks broadly similar to Figs.3,5, but increasing $S$
has the effect of deepening the minimum in $K(L)$ and hence allowing multiple
transitions where decoration by a spin one-half Ising spin would be
insufficient.

The Ising models discussed here have been living on an annealed ensemble of random
quadrangulations since the sum over graphs
in  equ.(\ref{sum}) is at the level of the partition function
and not the free energy. They therefore represent a very particular sort of annealed
geometrical disorder. A natural elaboration of the discussion here is
to consider the effects of similar, but quenched, geometrical disorder.
Although there have been no investigations of the case of quenched 
ensembles of $\phi^4$ graphs
or random quadrangulations, simulations of various spin models on 
quenched ensembles of $\phi^3$ graphs
or their dual triangulations strongly suggest that there is little, if any change,
in the critical couplings by comparison with the annealed ensemble \cite{JJ}.
The preceding
discussion of the effects of the decoration-iteration transformation could therefore
be carried over verbatim -- on quenched random quadrangulations
(allowing degenerate gluings) a competing anti-ferromagnetic and decorated
ferromagnetic bonds would be expected to give rise to multiple transitions.

Other models with quenched geometrical disorder can
also be subjected to the same treatment.
A Penrose tiling is just such a case, since
it is composed of rhombi and hence displays an anti-ferromagnetic
transition. 
The critical couplings here are $K_{crit} \sim \pm 0.41857..$
\cite{prz},
so as for the square lattice Ising model and the regular random quadrangulations
higher decoration, or higher spin decoration, would be required to 
force the system to display  multiple transitions. On the Penrose tiling
all of the transitions would be in the Onsager universality class. 

In summary, we have seen that geometrical disorder is no hindrance to
carrying out a decoration-iteration transformation and inducing
multiple transitions. The value of the critical coupling on
(annealed) random quadrangulations means that a single decoration is sufficient
to have an effect. It is likely that quenched random quadrangulations would
show a similar effect. It might be of some interest to explore the effect
of some of the other ``classic'' transformations, such as decoration-iteration
in field or star-triangle, 
for the Ising model coupled to 2D quantum gravity (i.e.
living on planar random graphs or their duals) since, at least implicitly,
the solution in field is available.  
 
%
\section*{Acknowledgements}
D.J. was partially supported by
the EC IHP network
``Discrete Random Geometries: From Solid State Physics to Quantum Gravity''
{\it HPRN-CT-1999-000161}. 
%
%
%
%

%
\clearpage \newpage
\begin{figure}[htb]
\vskip 20.0truecm
\includegraphics{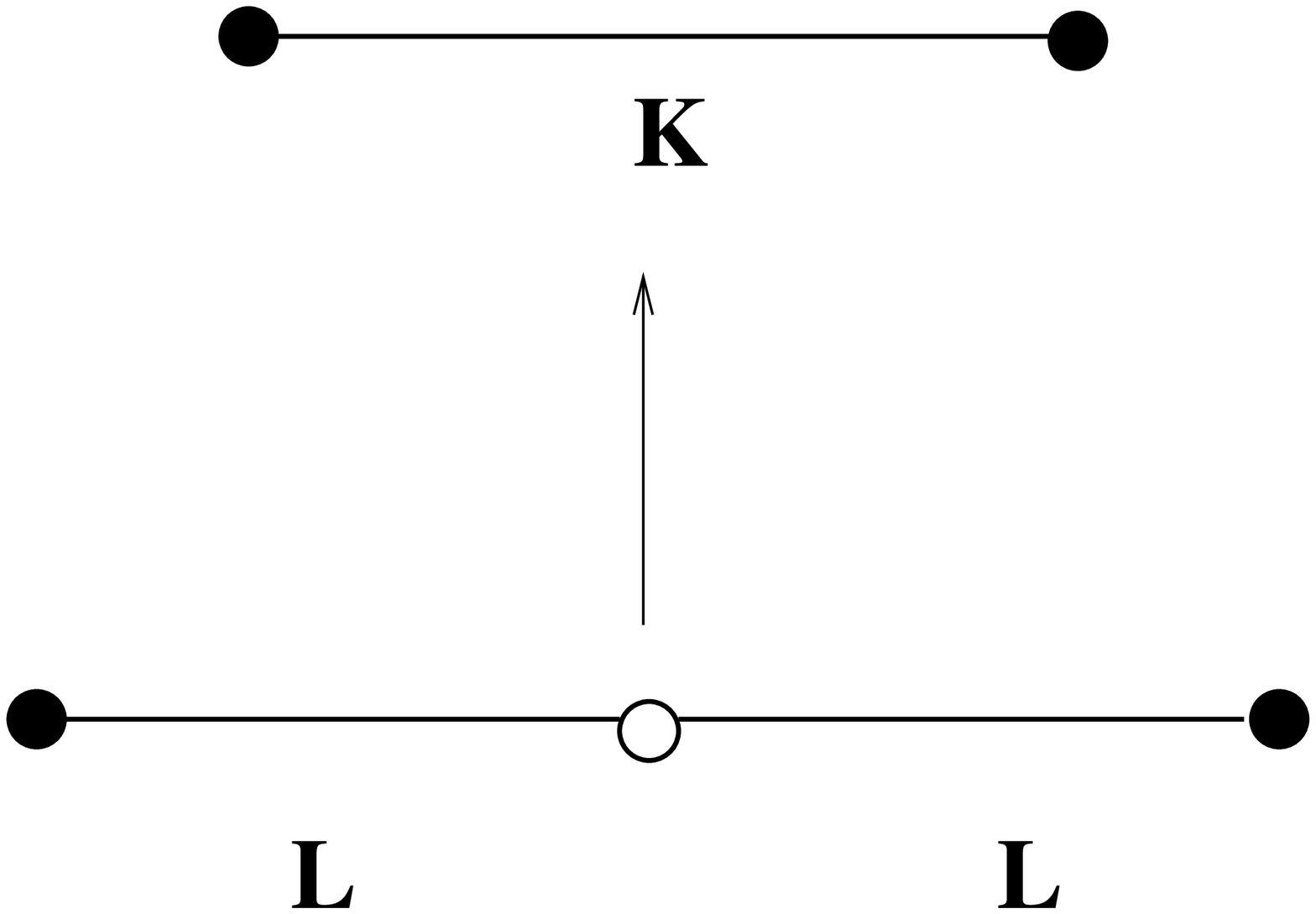}
\caption[]{\label{fig0} Summing over the central spin values
get an effective coupling $K$. 
}
\end{figure}
\clearpage \newpage
\begin{figure}[htb]
\vskip 20.0truecm
\includegraphics{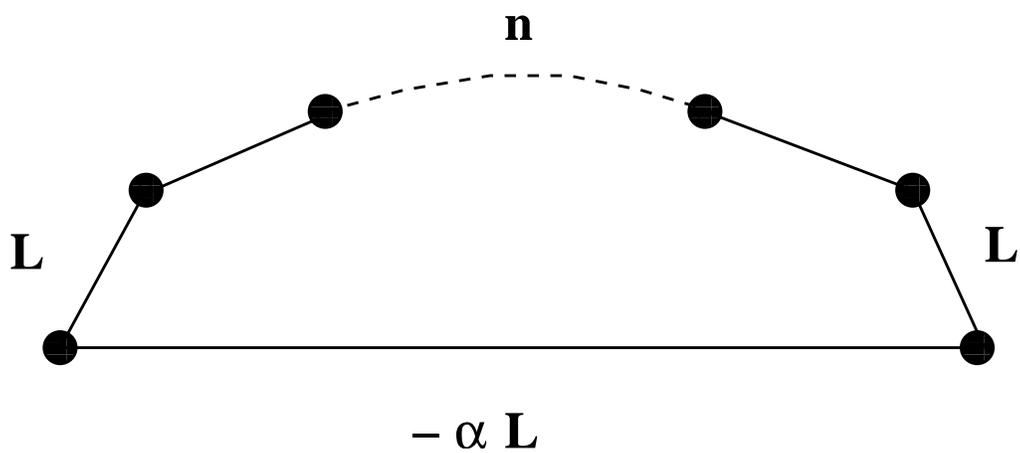}
\caption[]{\label{fig1} A direct antiferromagnetic bond of strength $-\alpha L$
decorated by $n$ spins coupled
by ferromagnetic bonds of strength $L$.
}
\end{figure}
\clearpage \newpage
\begin{figure}[htb]
\vskip 20.0truecm
\includegraphics{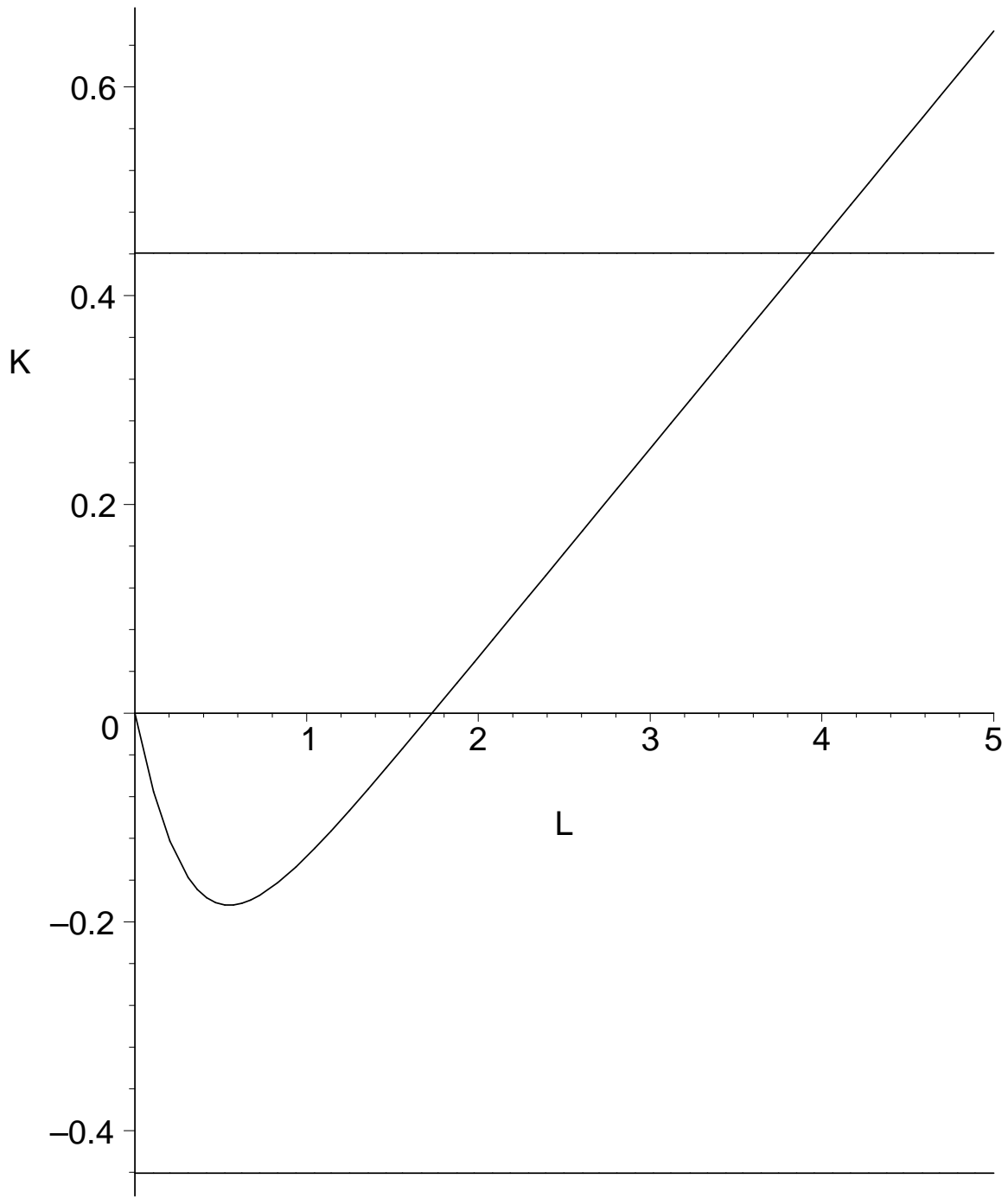}
\caption[]{\label{fig1a} 
For the square lattice ising model a singly-decorated bond as in Fig.1,
will {\it not} give multiple transition points since the curve $K(L)$
only crosses the ferromagnetic value $K = 0.44609..$. We have taken
$\alpha = 4/5$ in the figure.
}
\end{figure}
\clearpage \newpage
\begin{figure}[htb]
\vskip 20.0truecm
\includegraphics{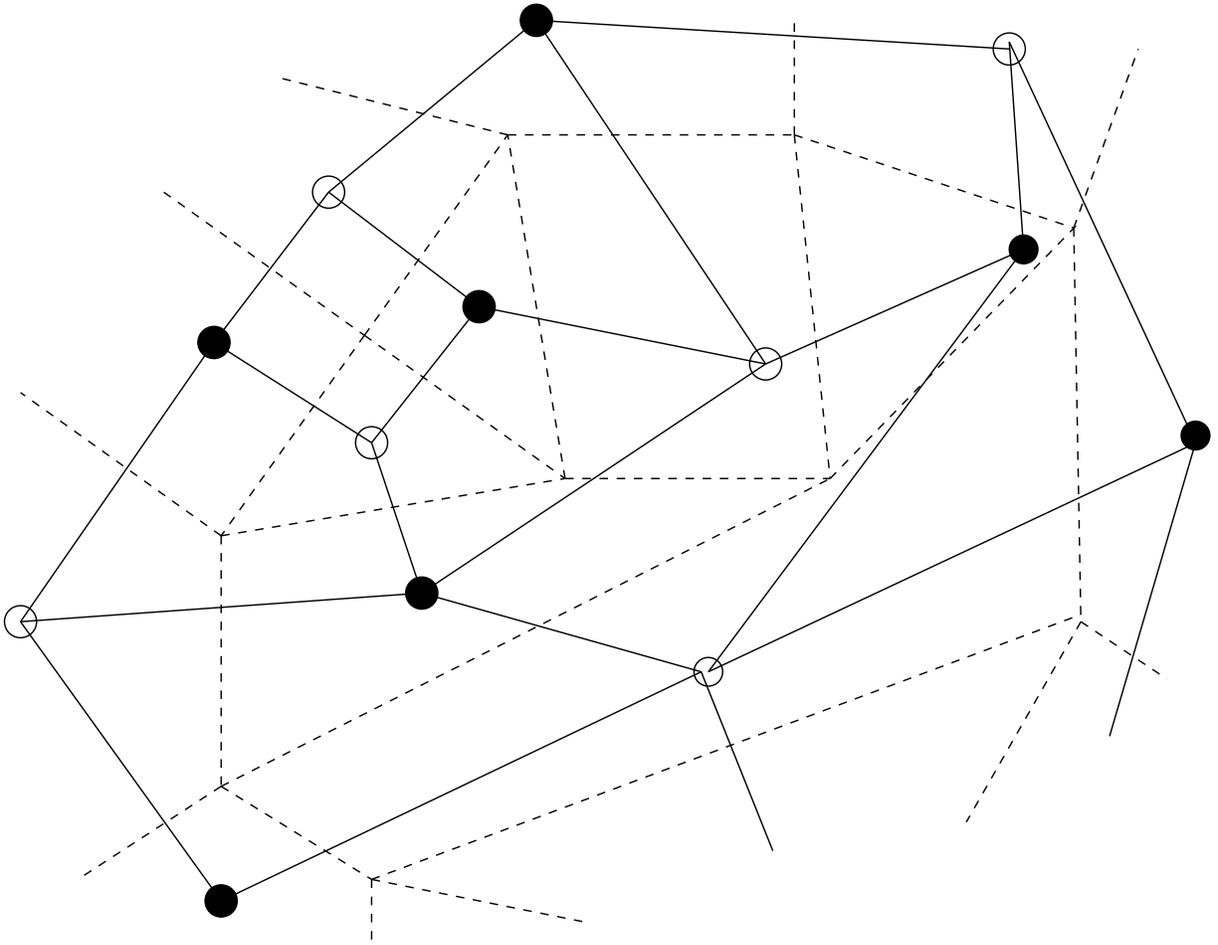}
\caption[]{\label{fig4} A section of a planar $\phi^4$ graph
and its dual quadrangulation. As the vertex shading makes clear
the  quadrangulation, although random, is still bipartite 
and hence will admit an antiferromagnetic transition.
}
\end{figure}
\clearpage \newpage
\begin{figure}[htb]
\vskip 20.0truecm
\includegraphics{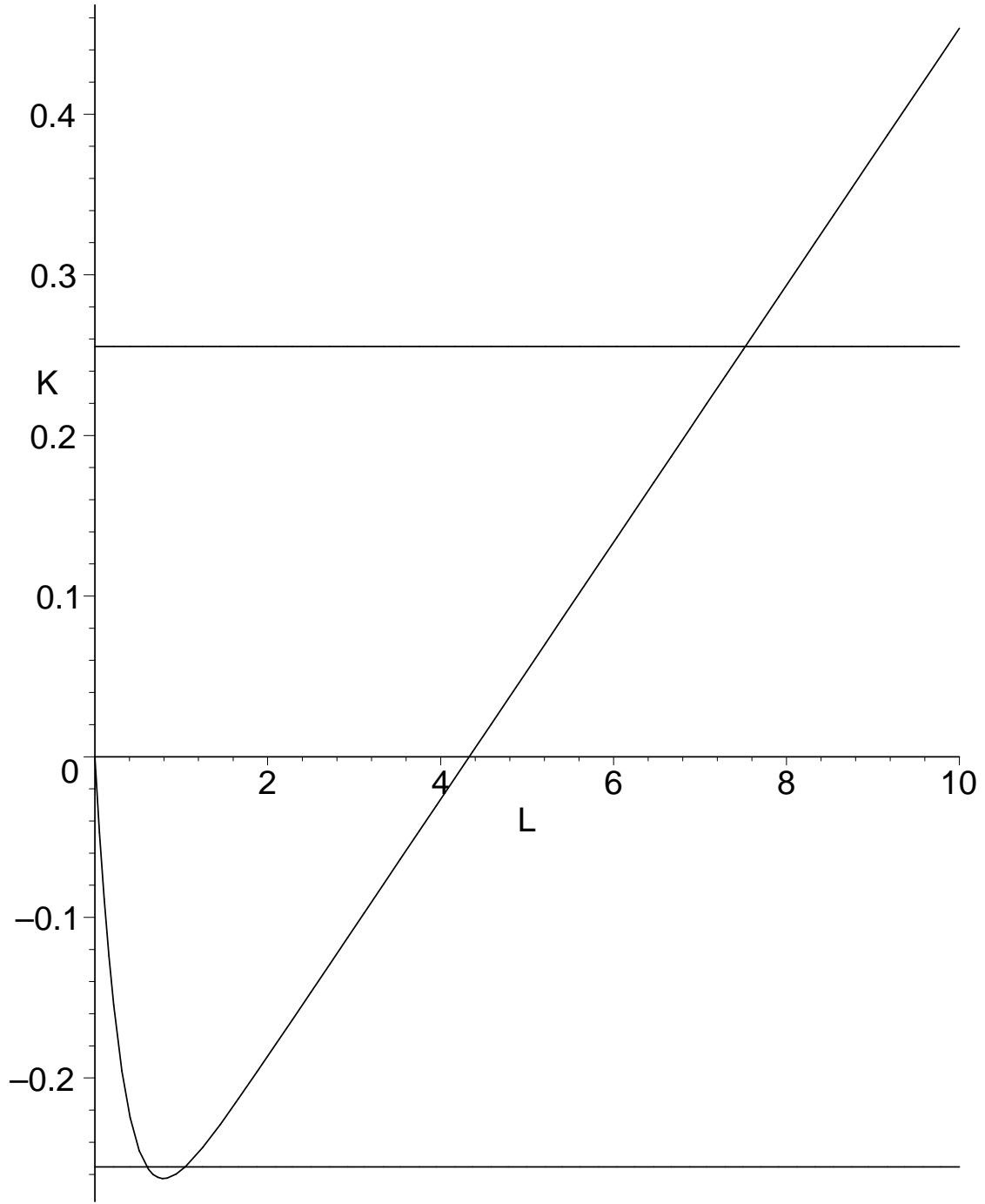}
\caption[]{\label{fig2} A plot of $K$ {\it vs} $L$ for
a singly decorated bond, $n=1$, $\alpha=0.92$. The values
of $K$ corresponding to the ferromagnetic $-(1/2) \log ( 3 / 5)
\sim 0.255412..$ and antiferromagnetic $-(1/2) \log ( 5 / 3)
\sim -0.255412..$ transitions 
on random quadrangulations are also shown. In this case
a singly decorated bond is sufficient to induce a triple
transition.
}
\end{figure}

\end{document}